\documentclass[sigconf]{acmart}
\renewcommand\footnotetextcopyrightpermission[1]{} 
\usepackage[normalem]{ulem}
\usepackage{booktabs}   
\usepackage{subfig}
\usepackage{graphicx}
\usepackage{tabularx}

\begin{document}

\title{Personal Volunteer Computing}         


\author{Erick Lavoie}
\affiliation{
  \institution{McGill University}            
}
\email{erick.lavoie@mail.mcgill.ca}          

\author{Laurie Hendren}
\affiliation{
  \institution{McGill University}            
}
\email{hendren@cs.mcgill.ca}          



\begin{abstract} 

We propose personal volunteer computing, a novel paradigm to encourage technical solutions that leverage personal devices, such as smartphones and laptops, for personal applications that require significant computations, such as animation rendering and image processing. The paradigm requires no investment in additional hardware, relying instead on devices that are already owned by users and their community, and favours simple tools that can be implemented part-time by a single developer. We show that samples of personal devices of today are competitive with a top-of-the-line laptop from two years ago. We also propose new directions to extend the paradigm.
\end{abstract}

\maketitle

\section{Introduction} 

The commercial success of computing has made available billions of personal devices. In the mobile industry alone, there has been more than a billion smartphones sold every year for the last 4 years~\cite{gartner2017cellphonesold,gartner2016cellphonesold,gartner2015cellphonesold,gartner2014cellphonesold}. In addition, millions of devices of yesteryear, including older laptops, tablets, and desktops are sitting at the bottom of drawers, in recycling centres and other warehouses. Since the average user obtains a new smartphone every 18-20 months~\cite{gartner2015replacementrate}, the next years are most likely to add billions more to the lot. These personal devices can potentially be used as a distributed computing infrastructure at no cost, should their current owners be willing to either actively lend CPU time or to donate the devices they no longer need. Given the availability of these devices, the resulting economic opportunity cannot be overlooked.

Personal devices will also be useful for longer than in previous times of rapid performance improvements. Hardware devices of today, excluding their battery, have a potential usable lifetime of at least a decade. Moreover, the slowing of Moore's law~\cite{theis2017end}, as well as the convergence of mobile computing performance with that of laptops and desktops of the recent past~\cite{herrera18webassembly} is likely to make their performance competitive for many years. 

We propose \textit{personal volunteer computing} as a novel distributed computing paradigm to leverage both opportunities. The paradigm aims at creating simple and effective designs for distributed systems that can quickly and easily tap into personal devices and at enabling volunteers from a user's community to contribute extra computing devices when needed. We target users with significant computing needs but limited resources available: potential developers that may only be available to develop and maintain new applications part-time with limited capital to acquire new hardware devices. This includes associations of citizens performing scientific activities in their spare time~\cite{publiclabs,zooniverse}, researchers in disciplines with limited funding available, or software developers and scientists in less industrialized countries. They could use personal volunteer computing tools for animation rendering, image processing, runtime verification, and many other applications.

This new personal volunteer computing paradigm is uniquely positioned, compared to existing paradigms, to answer the computing needs of developers with privacy needs and limited resources. This combination of characteristics had not explicitly been pursued by other major paradigms. Contrary to \textit{cloud computing}, computing resources can be used without financial transactions. Contrary to systems typically built for \textit{grid computing}, the computing resources are available to the general public with no administrative permissions. Compared to existing \textit{volunteer computing} tools, the tools are easier to deploy and require no dedicated hardware. Compared to \textit{edge computing}, which also leverages personal devices but require trust in an external platform operator, private data is only exposed to devices from trusted volunteers. Compared to other \textit{decentralized and peer-to-peer platforms}, the focus is on simple tools rather than a global platform, which removes the maintenance costs when not actively used.  

We show the applicability of personal volunteer computing today by measuring the combined performance we obtained on two samples of personal devices: the laptops and smartphones we have accumulated over the years, and the smartphones of our colleagues at work. 
We use Pando~\cite{lavoie2019pando,pando-repository}, a tool we built for personal volunteer computing applications, to show that both samples of personal devices, in their aggregate computing power, are competitive with a top-of-the-line laptop from two years ago.

Personal volunteer computing can further be extended in different directions, leading to interesting avenues of research. Newer approaches such as crowdsourcing~\cite{howe2006crowdsourcing}, in which participants take an active role in the tasks performed, or support for long-running computations, in which tasks and results may be exchanged during intermittent connections, would increase the number of compatible applications. The computations performed by devices could also be synchronized with the availability of energy; this would decrease the need for batteries, lower operating costs, and make the tools better compatible with intermittent energy sources.  

In the rest of this paper, we first articulate the context of personal volunteer computing compared to other popular paradigms of today to highlight its niche (Section~\ref{Section:MajorParadigms}). We then provide empirical evidence of the significant computing potential of older personal devices (Section~\ref{Section:ComputingPotential}). We finally identify future applications that are compatible with the paradigm and we articulate associated future research directions (Section~\ref{Section:FutureResearchDirections}).

\section{Major Socio-Technical Paradigms for Distributed Computing} 
\label{Section:MajorParadigms}

Distributed computing, through various paradigms, has developed according to social and economic factors that are often implicit behind the various technical contributions in the sub-disciplines. In this section, we make those factors explicit in order to identify a niche, which had not been explicitly articulated before, that we fill with \textit{personal volunteer computing} by focusing on personal applications, tools, devices, and social networks. This discussion shows that, compared to the other major paradigms of \textit{cloud}, \textit{grid}, \textit{volunteer}, \textit{edge}, and \textit{decentralized/p2p} computing, personal volunteer computing isß an original emerging socio-technical paradigm that opens up directions of research that would not necessarily be coherent with other paradigms. In the rest of this section, we briefly summarize the established paradigms and we end with a presentation of personal volunteer computing in Section~\ref{Section:PersonalVolunteerComputing}. Readers already familiar with existing paradigms may directly skip to Table~\ref{Table:Paradigms} for a high-level summary.

\subsection{Cloud Computing}
\label{Section:CloudComputing}

\textit{Cloud computing}~\cite{armbrust2009above,buyya2009cloud} has emerged ten years ago as a \textit{market service} that offers their \textit{customers} on-demand computing resources with no initial capital investment and quick scalability to match variations in resource usage. For many businesses, it offers (1) lower capital risks associated with over- or under-provisioning their hardware infrastructure to match the demand on their services and (2) enables economies of scale by sharing the same hardware resources among multiple users, therefore increasing resource utilization. Clouds accelerated the growth of startups into global platforms, AirBnB and Uber being notable examples.

The devices that power a cloud are provided by a \textit{single company}. The development and the management of the platform is \textit{funded by customers} using the cloud services directly, by renting the computing resources, or indirectly, by using online services that are implemented with them. Most often, cloud providers receive funding from both cases: Google provides ad-supported search services and the AppEngine~\cite{google-app-engine} platform; Amazon provides an online marketplace and the Amazon Elastic Cloud 2 (EC2)~\cite{amazon-ec2} platform. The exact number of devices cloud providers manage is considered a trade secret by some. Nonetheless, we estimate cloud providers may collectively manage in the order of \textit{millions} of devices.

The operating costs that have a direct impact on the profitability of operating a cloud (ex: hardware acquisition, hardware and software management, power and cooling energy requirements) incentivize their \textit{efficient} usage. Consequently, researchers develop strategies to build cloud infrastructure using commodity hardware, minimize resource consumption for given workloads, multiplex many concurrently running services on the same hardware to amortize the fixed costs of operation, and develop autonomic management strategies to minimize the involvement of humans. Companies also increasingly share designs both for hardware with the Open Compute Project~\cite{opencompute} and software with Open Stack~\cite{openstack} to lower the development and maintenance costs.  Additional challenges include the quality of service provided (ex: latency in provisioning resources, total amount of available computing power) and the accurate and efficient \textit{monitoring} of resource usage for \textit{billing} to ensure a customer only pays for what they use.

In their canonical form, clouds are limited in three ways. First, their billing infrastructure becomes a \textit{financial barrier} for individuals and organizations that do not have access to financial instruments, such as a bank account or a credit card. Second, their reliance on centrally-managed dedicated hardware entails a minimum price that may be inaccessible to many individuals and organizations. Third, their API requires access permissions which complicates their programming, in turn creating a  higher \textit{technical barrier}.

\subsection{Grid Computing}
\label{Section:GridComputing}

\textit{Grid computing}~\cite{foster2001anatomy,foster2008cloud} is an older but similar offering to cloud computing that makes computing resources \textit{belonging to different collaborating organizations} available through a unified service. Grid computing has been named by analogy to the way the electric grid was initially built. As grid computing developed, it was anticipated to also exist as a commercial offering, but was replaced by cloud computing. The grid approach survives today as a \textit{scientific utility} by providing computing resources to publicly funded organizations such as universities and research centres.
 
Grids are currently funded through the public spendings of \textit{governments} and offered to \textit{researchers} both to carry their research and train students in distributed computing.  We estimate the potential number of devices available through public organizations to be in the order of \textit{millions}, although specific grid projects have much lower offerings.  Grid5000~\cite{grid5000}, PlanetLab~\cite{chun2003planetlab} each currently boast offerings of about a \textit{thousand} devices.

The main challenge with grids is to create technical infrastructure that \textit{interoperates} with the various distinct administrative domains and organization policies that manage the computing resources while providing a \textit{unified interface} to researchers. 

Challenges in building grids are different than for clouds. There is no need for a billing infrastructure because in most cases both the users and computing infrastructure are paid from public spendings. Also, the focus is on collaboratively sharing the infrastructure between researchers rather than maximizing resource utilization because some research projects, such as performance studies, require a reserved access to the devices.

However, while grids are available to many public organizations, they are not available to the  \textit{general public}. Even for researchers, the administrative complexity in obtaining the necessary permissions may rule out many small-scale projects because the resulting gain is not worth the effort. Both issues raise \textit{administrative barriers}.

\subsection{Volunteer Computing}
\label{Section:VolunteerComputing}

\textit{Volunteer computing}~\cite{sarmenta2001volunteer,anderson2010overview} leverages the personal devices of volunteers from the \textit{general public} to perform computations.  It is organized around a \textit{commons} paradigm, both digital,\footnote{"\textit{The digital commons are defined as an information and knowledge resources that are collectively created and owned or shared between or among a community and that tend to be non-exclusivedible, that is, be (generally freely) available to third parties. Thus, they are oriented to favor use and reuse, rather than to exchange as a commodity. Additionally, the community of people building them can intervene in the governing of their interaction processes and of their shared resources.}" (Fuster~\cite{fuster2010governance})} by sharing tools between many independent research teams, and physical, by enabling volunteers to contribute their computing resources to many projects. Volunteers do not receive financial benefits for their contributions but may receive public recognition in the form of computation points. Computations performed with volunteer computing sometimes represent global issues, such as climate prediction~\cite{climateprediction} or drug discovery for cancer~\cite{folding@home}. Other times they simply capture the imagination of the general public, such as the search for extra-terrestrial intelligence (SETI@Home~\cite{seti@home}). 

The development of volunteer computing platforms has been funded by \textit{governments} through public research grants to provide \textit{researchers} with supercomputing capacities at a much lower cost. It has the potential to leverage \textit{billions} of personal devices although at the moment the current number of participating devices is in the order of a \textit{million}. At the time of writing, its flagship project, BOINC~\cite{boinc}, with 175,000 active volunteers managing 858,000 active computers, and a total combined computing power of 22.579 PetaFlops is one of the top five most powerful supercomputers in the world and has a fifth of the power of the most powerful one (Sunway TaihuLight), which boasts between 93 and 125 PetaFlops~\cite{top500}.
 
 The major typical challenges to the volunteer computing approach concern the \textit{variability} of capabilities of personal devices, the necessity to encourage and maintain volunteer \textit{engagement}, and the automatic handling of volunteer's \textit{unreliability}~\cite{anderson2006boincdesign,nov2010socialfactors}.
 
In contrast to cloud computing, an additional computation contribution does not cost the researchers, therefore efficiently using the hardware is less of an issue. Also, financial transactions are replaced with computation points that are issued after successful work has been performed, which lowers the \textit{financial barrier} to obtaining access to the computing devices. In contrast to grid computing, there are \textit{fewer administrative barriers} to deploy the tools: a research team may buy its own server and use the tools freely to request support from the general public.  Its major advantage compared to both clouds and grids, is that the majority of the costs are supported by volunteers, which covers the acquisition, the operation, and the maintenance of the computing devices.  However, it typically has a higher communication latency and more limited bandwidth available, making it better applicable to compute-intensive tasks with low communication requirements. 
 
 With 4 billion estimated Internet users~\cite{internet-stats-2017}, the current number of volunteers represents less than 0.005\% of humanity. From a technical perspective, the number of smartphones that were sold in 2015 and 2016 is more than 3000 times the number of active computers in volunteer computing projects~\cite{mobile-stats-2018}.  The approach has therefore not yet reached its full potential. We believe the complexity of the BIONC tools, that have been designed for researchers and large-scale projects, as well as the costs in acquiring and maintaining dedicated servers to run the them are remaining \textit{technical} and \textit{financial} barriers that slow the widespread adoption of the paradigm.

\begin{table*}[htbp]
\centering
\begin{tabularx}{\textwidth}{l l l l l l l}
\toprule
        & \textbf{Cloud} & \textbf{Grid} & \textbf{Volunteer} & \textbf{Edge/Gray} & \textbf{Decentralized} & \textbf{Personal}\\
        & &  &  & & \textbf{/P2P} & \textbf{Volunteer}\\
\midrule

\textbf{Motivation}            & Lower infrastr.  & Lower infrastr.       & Lower operation     & Lower operation & Resilience            & Lowest hardware \\
                                         & costs \& risks        & costs                & costs                       & costs                  & and sharing             & and op. costs         \\
\textbf{Paradigm}             & Market service     & Scientific utility  & Commons              & Market service    & -                          & Commons \\
\textbf{Ress. providers}    & Single company & Multiple org.       & General public       & End users            & -                           & Friends,family \\
\textbf{Target users}         & Customers         & Researchers       & Researchers          & Customers          & -                           & General public \\
\textbf{Funders}                & Customers         & Governments     & Governments         & Customers           & -                          & General public? \\  
\textbf{Nb of devices}        & Millions              & Millions              & Billions                   & Billions                   & Billions               &  Billions \\
\textbf{Challenges}           & Efficiency,         & Interoperability,   & Variability,               & Application            & Fault-tolerance,  & Simplicity, \\
                                         & monitoring,       & unified interface,  & engagement,          & partitioning            &  trust,                 & portability, \\
                                         &  billing              &  sharing                & unreliability             &                              & consensus         & scalability \\
\textbf{Trusted Parties}    & Platform op.      & Platform op.         & Tool op.                  &  Platform op.         & P2P algorithms   & Friends, family \\
\textbf{Design}                 & Global platform & Global platform & Persistent Tool & Global platform & Global platform & Transient Tool \\
\textbf{Coordination}        & Centralized & Centralized & Centralized & Centralized & Distributed & Centralized \\
                                         &                    &                    & (disjoint)      &                   &                         & (disjoint)     \\

\textbf{Coordinators(s)}    & Dedicated & Dedicated  & Dedicated & Dedicated & All devices & User device \\
                                         & servers     & servers       & server       & server       &                    &  \\
\bottomrule
\end{tabularx}
\caption{Comparison summary between the major distributed computing paradigms and personal volunteer computing.}
\label{Table:Paradigms}
\end{table*}

\subsection{Edge and Gray Computing}
\label{Section:EmergingCloudRefinements}

\textit{Edge} and \textit{gray} computing are based on a \textit{cloud}-hosted platform, and therefore inherit much of the same characteristics. \textit{Edge computing}~\cite{shi2016edge} performs computation tasks on the devices that directly interface with the real world, such as mobile phones and sensor networks. \textit{Gray computing}~\cite{pan2017thesis,pan2017gray} does the same in web browsers by offloading tasks to visitors of websites. In both cases the motivation is to provide better quality of service with lower latency and to lower the operation costs of cloud-hosted platforms. 

Edge and gray computing are similar to volunteer computing by their reliance on external resource providers on which operation costs are transferred, end-users in the former, and volunteers in the latter. They can both benefit by the large increase in number and performance of personal devices of the last years. However, the two approaches are quite different in all other aspects including the \textit{intention} of participants to contribute or not: volunteers intentionally choose to contribute to project they care about while end-users of edge computing platforms may unknowingly contribute resources to their operations. Moreover, as for cloud computing, end-users of edge computing platforms also have to \textit{trust the operators} that they will respect the privacy of their data and that the implicit computations performed on their device will not degrade the quality of service provided by their device.

\subsection{Decentralized Approaches}
\label{Section:DecentralizedDistributedComputing}

 \textit{Decentralized} and \textit{peer-to-peer} (P2P) approaches, contrary to all other paradigms, use personal devices both for the execution of tasks and their \textit{coordination}. This \textit{distributes} the usual responsibilities of servers in a central location to \textit{all the participating devices} in the network using \textit{peer-to-peer algorithms}. In turn, it makes the system more resilient to the failures of \textit{coordinators} by eliminating their privileged position.

The motivation of decentralized approaches is typically to increase the resilience of services by tolerating additional failure modes and spreading the loads differently. They are not tied to specific organization paradigms: some are used to implement decentralized storage~\cite{ipfs} and computation~\cite{iexec} through \textit{market services} supported by crypto-currencies while other are used to exchange files in \textit{commons} directly between users~\cite{bittorrent}.

Compared to cloud, grid, and volunteer computing, decentralized approaches remove the required trust from the \textit{operators of the platform/tool} and the \textit{servers} used. However, they usually still provide a \textit{globally shared platform} and accordingly maintain global structured overlays~\cite{therning2005jalapeno,abdennadher2005towards,nandy2005thesis,harrison2008thesis,kim2009thesis,kim2014scalable,wilson2015architecture,rosen2016thesis,dias2018browser} with corresponding  \textit{maintenance} and \textit{complexity} challenges. When applied to volunteer computing, maintaining the platform while it is not actively used puts pressure on volunteers to keep it running. This costs  energy, time, and attention, but provides no clear benefit. Moreover, the complexity in developing and maintaining such platforms requires dedicated specialists and ongoing recurrent resources.

Compared to decentralized approaches, volunteer computing tools are \textit{centralized} in the sense that the coordination of computations is performed on a \textit{dedicated server}. However, contrary to decentralized platforms, different users create \textit{disjoint networks}. This greatly simplifies the implementation of coordinators while providing independence from the failures of \textit{other users}.

\subsection{Personal Volunteer Computing}
\label{Section:PersonalVolunteerComputing}

We propose personal volunteer computing as a new volunteer computing approach that follows its \textit{commons} paradigm but focuses on the personal computation needs of programmers from the \textit{general public} for applications of personal or community interest. The user starts a computation on their personal device, then spreads the computations on other of their devices, and if the task still requires more computing power, they ask their friends, family, and colleagues to participate with additional devices. Its key opportunity is that it can provide distributed computing infrastructure without additional investment in hardware and at low operating costs by using participant devices for both coordination and computation with a simpler implementation than typical decentralized approaches. The main limitation of the approach is that availability of other volunteers' devices is not guaranteed because they are not integrated in an online platform; the devices instead join for punctual needs after explicit requests. The exact characteristics and performance of participating devices shall therefore be variable. Nonetheless, a user is likely to remember whom contributed the most computing power and will most likely invite previous volunteers with particularly powerful machines in priority for subsequent tasks.

The users we target are varied. They include but are not limited to scientists in research teams with low but significant computation needs,  individuals in developing countries with a personal smartphone and no access to other alternatives, amateurs doing science as a hobby, etc. The approach leverages a user's  \textit{trust} in the \textit{friends, family, and collegues} they are asking for help (their \textit{personal social network}) for two reasons: (1) it incentivizes more contributions since volunteers are more likely to contribute computing power to help someone they personally know, and (2) volunteers that are known by the user are less likely to intentionally provide invalid results since this would have social consequences to them once detected.  So far, our work has been funded from research grants from governments but its potentially wide applicability could encourage the \textit{general public to directly fund it for its own needs}. This approach could potentially use the \textit{billions} of  available personal devices~\cite{mobile-stats-2018}, but rather than unifying all devices in a single platform, devices assemble in temporary networks around potentially thousands of independent projects.

The main challenges for personal volunteer computing derive from the wide diversity of programming environments and software/hardware combinations to support, the more limited time available to learn and deploy the tools for small projects because they are often done part-time as part of other projects, as well as the currently limited capital available for its growth. It is therefore significantly more important than for other approaches that the tools remain \textit{simple} to use and to deploy to provide quick gains with low efforts. It is also important that the tools are quickly \textit{portable} to many environments, current and future, by being simple to implement. Finally, the tools also need to \textit{scale} to all the devices of the personal social network of its user to maximize their benefits.

Compared to cloud and grid computing, personal volunteer computing removes their \textit{financial} and \textit{administrative} barriers. Compared to volunteer computing, it drastically lowers its \textit{technical} barriers and removes its \textit{financial} barriers by using one of the \textit{user's devices} for coordination. In contrast to decentralized approaches, personal volunteer computing leverages the existing \textit{mutual trust between friends and family}. To recruit volunteers, social interactions, possibly through existing social platforms, are used instead of maintaining separate decentralized services. Both choices greatly \textit{reduce the complexity} of the infrastructure needed so that the \textit{tool} can be maintained by a single developer in their spare time.  

Personal volunteer computing, similar to volunteer computing, lends itself naturally to compute-bound applications with many independent tasks. While the computing needs of the projects targeted are smaller than for volunteer computing, the smaller network latency between local devices opens an opportunity for distributed applications that require more communication between computing nodes, extending the range of applications that can possibly be targeted.

\section{Computing Potential of Personal Devices} 
\label{Section:ComputingPotential}

For personal volunteer computing to be adopted, personal devices need to provide sufficient computing power to be useful. In this section, we show this is already the case by measuring the collective performance on CPU-bound applications that can be achieved using two samples of personal devices: a collection of personal laptops and smartphones we have accumulated over the years at home, and the smartphones of our friends at work. 

The experiments have been performed with Pando~\cite{pando-repository}, a new tool for distributing JavaScript computations on personal devices. It essentially provides a streaming map operation, which applies a function on every value of the stream and returns the results in order. The actual processing happens on participating devices in parallel. New devices may join anytime during computation simply by opening a URL in their browser and will obtain inputs to process and communicate back results through WebSocket~\cite{websocket} or WebRTC~\cite{webrtc}, depending on availability. Devices may also quit at any time without affecting the results; failed computations are transparently resubmitted to remaining devices. The current version has been optimized for throughput by avoiding redundancy when processing values and ensuring faster devices receive more values to process. We used version 0.17.2 of Pando for the tests. The design and implementation of Pando are covered in more detail in a separate publication~\cite{lavoie2019pando}.

\subsection{Personal Devices Experiments}

We have tested six applications that all use Pando to distribute the core and most expensive part of their computations. \textit{Collatz} implements the Collatz Conjecture~\cite{longest-collatz}, that has been made popular with the BOINC volunteer computing platform~\cite{anderson2004boinc}; our JavaScript implementation uses a Bignumber library to perform the recursive steps. \textit{Crypto-Mining} searches for a nounce  whose value, when combined with a block provided in input, will result in a hashed value with a certain number of leading zeros, similar to the proof-of-work algorithm of Bitcoin~\cite{nakamoto2008bitcoin}. \textit{Random-testing} simulates the behaviour of the StreamLender abstraction, at the heart of Pando, on random interleavings of concurrent processes to find examples in which execution properties are violated to ensure they never happen in practice. \textit{Animation-rendering} renders individual frames of a synthetic scene by applying a raytracer algorithm on each of them and then assembles the result in a gif animation. \textit{Image-processing} applies a blur on satellite images of the Landsat-8 open dataset~\cite{landsat8}. \textit{MLAgent-Training} trains an agent in a simulated environment over a sequence of steps using reinforcement learning~\cite{convnetjs}. All applications are CPU-bound and we hide transmission delays by sending values to process in batches of two, this way network delays for one value happen while the other is processed. We used the version of benchmarks at commit \texttt{12164ee69b} of the pando-handbook~\cite{pando-handbook}.

Table~\ref{Table:ThroughputPersonalDevices} shows the combination of devices we gathered from those we have accumulated over the years. The oldest is the iPhone 4S (2 cores 1.0 Ghz ARM 32-bit), released in 2011, and the two newest are the  iPhone SE (2 cores 1.85 Ghz ARMv8 64-bit), released in 2016, and the Macbook Pro 2016 (4 cores i5 2.9 Ghz x86 64-bit). In between, we also have the  Novena~\cite{novena}, a linux laptop based on a Freescale iMX6 CPU (4 cores 1.2 Ghz ARMv7 32-bit) produced in a small batch in 2015, an Asus Windows laptop based on a Pentium N3540 (4 cores 2.16 Ghz x86 64-bit) processor, and a Macbook Air mid-2011 (2 cores i7 1.8 Ghz x86 64-bit). We used Firefox  (64.0 on x86 and 60.3.0 ESR on ARM) on laptops for consistency and because it is the fastest on numerical benchmarks~\cite{herrera18webassembly}; on the iPhones we used Safari. 

We noticed that the number of concurrent browser tabs that provided the maximum performance was less than the number of cores of many devices, possibly because some shared resources of the CPUs were saturated or because the OS or the browser reserved other cores for different services. We therefore chose the minimum number of cores that provided the maximum performance, which we mention beside the device name in Table~\ref{Table:ThroughputPersonalDevices}. The performance when using a single core was roughly equal to the ratio of the throughput obtained divided by the number of cores mentioned. We also reserve one core on the MacBook Air 2011 to execute Pando's master process which coordinates communication with other devices, leaving the other for computations.

A few results are worth discussing. First, the performance of the iPhone 4S was too low on some benchmarks to be included. On the others we noticed that the iPhone SE brings a significant performance improvement, between 3x and 21x. This shows that not all older phones may provide a significant contribution on modern tasks. Second, the iPhone 4S and the Macbook Air 2011 are of the same generation, similar to the iPhone SE and the Macbook Pro. The performance gap between each pair, when taking the performance on a single core, has significantly reduced; it was between 3.3x and 14x in 2011 and dropped to between 1.3x and 2.1x in 2016. Note that on the image processing application, the Macbook Pro is surprisingly slower; using Safari on the Macbook Pro instead of Firefox makes it faster again, the difference can therefore be attributed to the difference in optimizations performed by browsers. Third, combining all other devices provides a performance level comparable to that of the Macbook Pro, which means that we could at least double the overall throughput of the applications by leveraging other devices we have access to, making them quite useful.

\begin{table*}[htbp]
\begin{tabular}{l r r r r r r r r r r r r}
                                                       & \textbf{Collatz}  &            & \textbf{Crypto-}   &           & \textbf{Rand.-} &      & \textbf{Anim.-} &             &  \textbf{Image-} &           & \textbf{MLAgent-} &  \\
                                                       &                           &            & \textbf{Mining}   &            & \textbf{Test.} &         & \textbf{Render.} &           & \textbf{Process.} &          & \textbf{Training} & \\
                                                        & \textit{Bignum/s} & \textbf{\%}   & \textit{Hashes/s} &  \textbf{\%} & \textit{Tests/s} &\textbf{\%}  & \textit{Frames/s}  & \textbf{\%} & \textit{Images/s} & \textbf{\%} & \textit{Steps/s} & \textbf{\%}  \\
\hline
\textbf{Device \textit{(cores)}} & & & & & & \\


iPhone 4S \textit{(1)}                        &      15.55       & 0.9        &    13,951      &  3.5         &      54.22    &   1.5       &         ---       &             ---  &          ---     &       ---    &         22.02  &   4.2 \\
Novena \textit{(2)}                            &     63.56     &    3.8        &  16,326    &   4.2              &   150.46   &    4.1       &         0.34  &          3.2 &         0.03   &          5.2  &        51.76  &   10.0  \\
Asus Laptop  \textit{(3)}                   &    254.08    &   15.1       &  59,877   &      15.2        &   617.40    &    16.7     &          1.88  &         17.6  &        0.08   &        15.9  &      112.23  & 21.6 \\
\textit{MBAir 2011} \textit{(1)}          &    218.92    &    13.0      &  56,906    &     14.5        &    551.18   &    14.9    &          1.47  &          13.7 &        0.04   &          7.5  &          72.88  & 14.0 \\
iPhone SE \textit{(1)}                       &   314.86    &    18.7      &  46,849   &       11.9       &       498.65  &    13.5     &         1.69   &        15.8  &       0.23    &          44.3  &       63.81 & 12.3 \\
MBPro 2016 \textit{(2)}                   &    814.48    &    48.4      &  199,917    &       50.8        &  1816.23  &     49.3     &          5.33  &        49.8  &        0.14  &          27.1   &     197.17  & 37.9 \\

\hline
\textbf{All}                                       &     1681.45   &        100.0  &  393,826 &        100.0 &  3688.15    &         100.0 &        10.70    &        100.0 &       0.53    &                     100.0 &  519.86       &         100.0 \\
\end{tabular}
\caption{\label{Table:ThroughputPersonalDevices} Average throughput for CPU-bound streaming applications using a combination of personal devices.}
\end{table*}

\subsection{Smartphones Experiments}

Table~\ref{Table:ThroughputCellPhones} shows a repetition of the random-testing experiment, this time with the smartphones of our colleagues at work, which arguably represent an interesting sample of those available. We only experimented on a single application, since there was a bound to our colleagues' interest in witnessing their batteries being drained. The exact specification of each device is rather tedious to list and of limited interest since the whole experiment would be rather hard to replicate. There are still a few points worth mentioning.

\begin{table}[htbp]
\centering
\begin{tabular}{l r r}
\textbf{Random-Testing} & & \\
 & \textit{Tests/s} & \textit{\%} \\
 \hline
 \textbf{Device} & & \\
iPhone SE & 443.46 & 18.70 \\
Huawei P10 lite 2017 & 364.99 & 15.39 \\
Samsung Galaxy S7 & 304.64 & 12.84 \\
Xiaomi redmi note 6 pro & 291.03 & 12.27 \\
LG G6 H870 2017 & 260.17 & 10.97 \\
Lenovo P2a42 2016 & 171.26 & 7.22 \\
Wileyfox Storm 2016 & 128.89 & 5.43 \\
Honor & 125.29 & 5.28 \\
Zenfone 3 & 100.58 & 4.24 \\
Samsung A3 2016 & 90.58 & 3.82 \\
Zenfone 2 & 55.81 & 2.35 \\
Huawei P10 lite 2017 (2) & 35.13 & 1.48 \\
\hline
\textbf{All} & 2371.82 & 100.00 \\
\end{tabular}
\caption{\label{Table:ThroughputCellPhones} Average throughput of smartphones volunteered by colleagues.}
\end{table}

First, the range of performance is significant, the slowest device, the Zenfone 2, is 8 times slower than the fastest of the lot, the iPhone SE. Second, it may be possible that some of them had been using energy saving modes, the iPhone SE was connected over a usb cable while the other were all running from their batteries. This is certainly the case for the second Huawei phone, which locked during the experiment and went into low power mode, explaining the 10x difference with the other identical device. Third, the overall performance of all devices combined is higher than that of the Macbook Pro, showing that asking your colleagues for help may be a valid substitute for a faster machine on some applications. And finally, the implementation of the application used only a single core on each device, it may be possible to reach a factor of 2-4 better performance in the future by leveraging parallel libraries or WebWorkers~\cite{webworkers}.

The last experiment shows that while older smartphones, such an iPhone 4S, may contribute an insignificant amount of computing power, the combined computing power of a dozen more recent smartphones can outperform a top-of-line laptop of only two years ago. Moreover, the first experiment results provide additional empirical evidence for the decreasing performance gap between smartphones and laptop computers, opening the door for using them for significant computations in the future, not only alone but also in combination with others.

\section{Future Research Directions} 
\label{Section:FutureResearchDirections}

In this section, we envision, from the \textit{personal volunteer computing} paradigm, a larger scope of application than has been shown in the previous section and follow with the research directions they open up. This should help establish a research community effort and provide interesting directions of investigation.

As a starting point, we take as a core assumption that a growing number of sectors of our society are currently reaching limits to growth~\cite{meadows1974limits}, be them available governmental funding for less popular research fields, public spendings to support various social services, household surpluses available to support various charities and citizen initiatives, etc. This does not preclude some other sectors from still experiencing dramatic growth and improvement rates. Nonetheless, to keep computing technologies relevant for sectors with less or no economic growth, it is important that the systems designed for them are as affordable as possible. The key advantage of personal volunteer computing to that end is that it can provide computing services without additional investment in hardware and could synchronize with intermittent and free energy sources to provide extremely low operating costs.

We first review concrete potential future applications from existing citizen science initiatives. We then articulate a set of design principles that would be consistent with providing computing services without small or no investment in hardware and low operating costs. We then sketch what potential research topics may be pursued within different sub-fields of computer science.

\subsection{Potential Applications}

The unifying theme behind all the applications we envision is that they \textit{increase capabilities at the community level by using that same community's resources} rather than distant computing infrastructure and associated supporting resources. We start from applications that may directly contribute to ongoing initiatives and then generalize abstract \textit{dimensions} that may guide future investigations and associated technical solutions.

One potential direction with potential short-term gains is to support existing citizen science initiatives. Some Public Lab~\cite{publiclabs} projects rely on near-infrared imagery to determine plant health~\cite{Infragram,ndvi} and rely on software processing pipelines~\cite{image-sequencer}. Pictures are often processed at home, due to the amount of processing required. Using Pando would enable that processing to happen in the field with the volunteers' smartphones. Another example, Zooniverse~\cite{zooniverse}, leverages the abilities of volunteers to perform classification, pattern-matching, annotation, and transcription tasks that may provide useful data for researcher as is, or after training with machine learning algorithms. Pando in this case could make the effort more social by coordinating the efforts of volunteers working in the same room on their smartphones to perform the different tasks. In fact, we already implemented a similar example to perform collaborative filtering of a stream of interesting Arxiv paper abstracts~\cite{pando-handbook}.

The applications we used in Section~\ref{Section:ComputingPotential} all used the devices \textit{for automatic processing}, \textit{synchronously to complete tasks faster}, and \textit{in the vicinity of its user}. Each choice is one possibility along a different dimension. We briefly sketch other possibilities along the same dimensions.

\textbf{Nature of Computing}. \textit{Automatic computations} are performed strictly on machines. \textit{Human computations} require the input from a human to make a decision, identify features, classify elements, or process information, such as those performed by Zooniverse volunteers and other forms of crowdsourcing~\cite{howe2006crowdsourcing}. In between, \textit{hybrid computations} may blend both for a better result. An exciting prospect for personal volunteer computing is to use data generated by a community and train machine learning algorithms to act as a collective memory specific to that community. For example, a foraging application, that could help identify and track useful plants and mushrooms and optimized for a local region, could be quite valuable to spread the skills quickly while preventing the appropriation of the knowledge by external parties.

\textbf{Locality in time}. \textit{Synchronous processing} has all computing resources working at the same time to complete a task as quickly as possible. \textit{Asynchronous processing} decouples the work performed by computing resources, which may instead contribute when they are most available. This would enable, for example, processing infrastructure powered by renewable energy in which the nodes work when there is sufficient energy available but otherwise power down. This could work well in combination with an asynchronous messaging infrastructure such as Secure Scuttlebutt~\cite{secure-scuttlebutt}.

\textbf{Locality in space}. \textit{Colocated computations} happen with all computing resources in a close physical space. \textit{Widely distributed computations} happen with computing resources spread over a larger area and potentially connected by routing infrastructure, such as the Internet. The latter may happen for online communities of interests. We can imagine, for example, a community-specific archival service, similar to Archive.org, that would only keep track of resources that were linked by community discussions related to specific interests. It would use the computing power and storage of participants to archive the content at links, build indexes of resources, and spread the load among members sharing the same interests.

All the applications proposed previously address a community's needs \textit{using the computing resources it already owns}. We believe there are many more interesting applications that can be built along the same lines and these have the potential to be several orders of magnitude more affordable in infrastructure investment and energy usage than current alternatives.

\subsection{Additional Design Principles}

We propose the following additional design principles for a new generation of personal volunteer computing systems.

\textbf{Local first.} Instantly available high-bandwidth communication and scalable computing power and storage planet-wide require significant infrastructure investment and uses enough energy to power small countries. In contrast, systems keeping information and computation local for services that may be offered without global connectivity will make the services more resilient to potential outages and energy shortages, and will bring down energy usage in intermediate routing nodes and data centres by several orders of magnitude.

\textbf{Leverage humans and their communities first, then augment their capabilities.} Volunteers can achieve gigabytes/s by carrying portable hard-drives; they can quickly provide access to dozens of CPUs in a few minutes by lending their devices' capabilities; they can help organize information and maintain devices in working order, and they can build supporting infrastructure for providing energy and communication.  Moreover, effective human communities have a high inherent level of trust and solidarity that can help simplify the designs of systems. The total amount of voluntary work volunteers can bring to keep their community functioning is a large resource to leverage. Volunteers may still benefit from automation to decrease the most tedious aspects of their work, in this case the infrastructure augments their capabilities rather than replacing them.

\textbf{Optimize for energy use and local energy sources.} Lowering the energy usage drastically reduces the operating costs. There are many potential optimization opportunities for applications that are not latency or throughput critical. This may allow, for example, to create systems whose computation activities follow the natural cycles of energy availability from solar panels or wind turbines. This would in turn lower the need for expensive energy storage, or backup infrastructure powered by non-renewable energy, therefore further lowering the infrastructure and maintenance costs.

Taken together, these principles really represent an exciting new paradigm of research that considers both unexploited opportunities of today and the hard constraints on energy and material usage that may appear in a near future for many sectors of our society.

\subsection{Research Topics}

In this section, we consider the promising avenues to be explored in existing disciplines of computing and other related fields when designing new systems built along the previous principles.

\subsubsection{Programming Languages}

To enable a single, or a few developers, to maintain and extend the required software infrastructure favours a renewal of the tradition of self-hosted programming systems, such as Oberon~\cite{wirth2007modula}, Smalltalk~\cite{goldberg1983smalltalk}, and Forth~\cite{brodie2004thinking}. Newer efforts could aim at creating a full environment for community applications based around personal volunteer computing.

\subsubsection{New Algorithms}

We envision the following three new kinds of algorithms to be implemented for personal volunteer computing systems.

\textit{Social algorithms} would be partly or completely be performed by cooperative members of communities, while being augmented by capabilities of interactivity, long-term faithful storage, and automatic cryptographic validation brought by the computers.

\textit{Subjective algorithms} would index, process, and organize produced by local communities according to the interest of each user, from their point-of-view, without access to global information. This could provide similar services as current recommendation and personalization services of today without requiring a third-party to access private data.

\textit{Natural algorithms} would be long running and energy-aware so they behave in accordance to the natural cycles in which the computing systems are embedded.

\subsubsection{System Design}

For many older devices, the maintenance and support from the original manufacturer is discontinued earlier than the end of their useful lifetime. For example, Apple stops supporting updates for older devices six years after they have been released. As such there is a need for minimal operating systems that can be developed and maintained by a community inheriting legacy devices long after their manufacturer stopped their support, which could span multiple decades for some devices. The designs should also take energy availability into consideration for task scheduling, persistence, and performance management.

\subsubsection{Communication and Sensor Networks}

Old mobile phones may be used to design ad hoc mesh communication networks. The resulting system should be self-configuring with minimal expert knowledge from volunteers deploying them. New routing algorithms should also take into account energy availability and efficiency.

\subsubsection{Energy Engineering}

Design of small-scale energy storage and production in the 5-10W range that can be built with local, abundant, and inexpensive materials by volunteers. These could be based on various technologies, including sterling engines using water or oil for thermal storage, thermo-electric effects by combining different metals and heating them with the sun, small wind or water turbine build with salvaged electric motors, etc. These will help bring the operation costs close to zero.

The previous research directions, compared to the current trends in research, make smaller whole-system designs done by small teams viable again. There may therefore be valuable insights to dig back from the 80s and 90s literature and to refresh with the benefit of insights from the two to three decades that followed, including the growing importance of energy management.

\section{Conclusion} 

In this paper, we have articulated a novel paradigm for distributed computing, which we named personal volunteer computing, that: (1) leverages existing personal devices, such as smartphones and laptops, (2) encourages the development of system designs that can be implemented by a single developer part-time, and (3) applies to the development of personal applications with significant computation needs, such as animation rendering and image processing. Two relatively new factors concur to make it possible: there is a current abundance of personal devices with a usable lifetime significantly longer than their current replacement rate and the performance improvements between generations of devices have decreased to the point that combinations of older devices are competitive with newer. We corroborated this analysis by measuring the performance obtained on a sample of smartphones and laptops we have accumulated over the years, as well of smartphones of our friends and colleagues on a number of representative applications. Both samples have been shown to be competitive with a top-of-the-line laptop from two years ago. To foster further developments on personal volunteer computing, we finally sketched a larger scope of applications, additional design principles for creating new tools, as well as potential topics to research within disciplines inside and outside computer science. They hold a promise of affordable computing for developers with significant computing needs but limited resources available.

\section{Acknowledgements}                            
This material is based upon work supported by the National Science and Engineering Research Council (NSERC) of Canada, Fond de Recherche du Quebec -- Nature et Technologies (FRQNT).  Any opinions, findings, and conclusions or recommendations expressed in this material are those of the author and do not necessarily reflect the views of NSERC or FRQNT.
  
\bibliographystyle{ACM-Reference-Format}
\bibliography{ErickLavoie}

\end{document}